# Integrating Simulation Budget Management into Drum-Buffer-Rope: A Study on Parametrization and Reducing Computational Effort

Balwin Bokor[1], Wolfgang Seiringer[1] and Klaus Altendorfer[1]

[1]University of Applied Sciences Upper Austria,
Department for Production and Operation Management
Wehrgrabenstraße 1, 4400 Steyr, Austria
`{balwin.bokor, wolfgang.seiringer, klaus.altendorfer}`
`@fh-steyr.at`

**Abstract.** In manufacturing, a bottleneck workstation frequently emerges, complicating production planning and escalating costs. To address this, Drum-Buffer-Rope (DBR) is a widely recognized production planning and control method that focuses on centralizing the bottleneck workstation, thereby improving production system performance. Although DBR is primarily focused on creating a bottleneck schedule, the selection of planning parameters is crucial, as they significantly influence the scheduling process. Conducting a comprehensive full factorial enumeration to identify the ideal planning parameters requires substantial computational effort. Simulation Budget Management (SBM) offers an effective concept to reduce this effort by skipping less promising parameter combinations. This publication introduces a method for integrating SBM into multi-stage multi-item DBR planned and controlled production system with limited capacity, aimed at determining the optimal planning parameters. Furthermore, we conduct a simulation study to analyze the effects of different production system environments, i.e., varying levels of shop load and process uncertainty, on both the performance and parameterization of DBR and the efficacy of SBM. Our results show significant reduction in simulation budget for identifying optimal planning parameters compared to traditional full factorial enumeration.

## 1    Introduction

Production systems often face bottlenecks, described by Goldratt [1] in the Theory of Constraints as capacity constraint resources. Bottlenecks, where the slack between required and available capacity is minimal or nonexistent, play a pivotal role in determining a production system's overall throughput and significantly impact key performance indicators (KPIs) [2]. While bottlenecks in production systems can take various forms, including specific employees or machine tools [3], they are most commonly associated with particular workstations [4]. To ensure favorable KPIs, it is crucial to identify bottlenecks [5] and strategically manage bottleneck capacity, focusing on aligning production with customer demand [6]. Hence the entire production system must be subordinated to the bottleneck workstation to ensure optimal performance. This principal is emphasized by the widely recognized production planning and control method Drum-Buffer-Rope (DBR). The selection of planning parameters significantly influences the bottleneck schedule and, consequently, affects KPIs and costs [7].

Given that real-world production systems must operate in stochastic environments with intricate interdependencies, simulation emerges as a particularly effective approach for determining these planning parameters [8]. However, to attain statistically reliable results in a stochastic environment, a substantial number of replications, involving simulations with equal parameterization but varying random number perturbations, is essential [9]. The full factorial enumeration approach, despite its precision, is time-consuming and may even be impractical as combinatorial complexity rises, particularly when it involves a complete enumeration of the entire feasible solution space [10]. A simple and fast concept to reduce simulation time while maintaining solution quality is achieved through Simulation Budget Management (SBM). This concept minimizes computational effort by skipping non-promising replications, as demonstrated by Seiringer et al. [11] in the context of a material requirements planning (MRP) production system. Extending its integration into other production planning and control methods will validate the broad applicability of SBM. Therefore, this publication integrates the SBM concept into a DBR planned and controlled multi-item multi-stage make-to-order flow shop production system with limited capacity. A real-world scenario similar to our observed setup includes automotive assembly line and electronics manufacturing assembly lines, both organized as flow shops with critical bottlenecks at complex stations like engine installation and smartphone assembly.

The primary objective is to reduce computational efforts when determining optimal planning parameters for DBR through a simulation-based approach. From a scientific standpoint, the study explores whether the SBM concept yields the same parameterization as a full factorial enumeration. Additionally, it explores the impact of different levels of shop load and process uncertainty on the parameterization of DBR and the performance of the SBM concept. From an economic perspective, decision-makers are presented with a practical method to reduce computational complexity in simulation studies and insights into how shop load and process uncertainty affect optimal planning parameters of DBR and costs. Hence, the following research questions are addressed:

RQ1: How do different production system environments, i.e., changes in shop load and process uncertainty, influence the performance and optimal planning parameters of DBR?

RQ2: Do different production system environments, i.e., shop load and process uncertainty, influence the performance of SBM?

RQ3: What is the risk of overlooking the optimal solution when employing SBM concepts?

RQ4: What potential savings in simulation budget (effort) can be realized through the implementation of the SBM concept at DBR, and what constitutes the optimal SBM parameterization?

The rest of the article is organized as follows: Section 2 provides a brief introduction to the production planning and control method DBR and introduces the concept of SBM for achieving optimal solutions within limited simulation resources, along with a discussion of related strategies. In Section 3, we explore the production system, describe the abstracted simulation model, and offer insights into the numerical study. Section 4 presents the outcomes of our simulation study, along with a comprehensive examination and discussion of the findings. Finally, in Section 5, we outline the primary contributions of our research and set the stage for future research directions.

## 2      Related Work

In this section we provide an overview of the production planning and control method DBR, including a comprehensive description of the forward scheduling approach and the applied planning parameters. Subsequently, we outline the SBM concept, as well as other related strategies, aimed at minimizing computational effort in simulation studies.

### 2.1     Drum-Buffer-Rope

The production planning and control method DBR is introduced by Goldratt and Cox [2]. Its fundamental concept revolves around scheduling the bottleneck workstation, thereby aligning the entire production system with the throughput of this critical constraint. In the DBR framework, the gross requirements are batched into lot sizes by the Master Production Schedule (MPS) to generate production orders, often utilizing Periodic Order Quantity lot-sizing policy [12]. After establishing the lot size, the bottleneck must be scheduled. This scheduling effectively requires two main buffers, which act as time horizons analogous to planned lead times in MRP [13]. However, DBR distinguishes itself by setting these two planning parameters at a system-wide level, applying uniformly across all materials concerning a particular production path, which reduces the number of planning parameters in comparison to MRP [12]. The two planning parameters are the Shipping-Buffer, which functions to prevent delays in completion, and the CCR (capacity-constrained resource)-Buffer, aiming to protect the bottleneck from starvation [14].

In DBR framework, a critical decision is whether to choose forward or backward scheduling for the bottleneck workstation. Forward scheduling determines when production at the bottleneck will be completed, identifying tardiness through planned end dates that fall after due dates. Whereas, backward scheduling determines when production should start, identifying tardiness through planned start dates that are set in the past. Although both approaches have benefits, as discussed by Mleczko [15], this study focuses on forward scheduling, primarily due to its avoidance of planning start dates in the past.

For applying forward scheduling, let's define $I$ as the set of all production orders, where each order $i \in I$ is associated with a due date $d_i$. A scheduling window is established by identifying production orders whose due dates $d_i$, after subtracting the sum of the Shipping-Buffer $S$ and CCR-Buffer $C$, fall beyond the current time $t$. The scheduling is performed iteratively, triggered whenever a new production order is within the scheduling window or when a production order is completed at the bottleneck workstation. Consequently, the subset of production orders to be scheduled at time $t$, denoted as $O(t)$, can be formulated as:

$$O(t) = \{i \in I \mid d_i - (S+C) \leq t \wedge i \notin R(t)\} \qquad (1)$$

Here, $R(t)$ represents the set of production orders that have already been released for production. By excluding these released production orders from $O(t)$, a 'frozen zone' is created, effectively minimizing system 'nervousness' that arises from production orders with imminent due dates [16]. Once the set of production orders $O(t)$ for scheduling is established, they are sorted in ascending order according to their due dates $d_i$, thereby implementing an Earliest-Due-Date (EDD) approach for order release. When a production order is scheduled (at the moment when $i \in O(t)$), the earliest plan start date at the bottleneck workstation $a_i$ is calculated by adding the CCR-Buffer $C$ to the current time $t$. Given this earliest plan start date at the bottleneck workstation $a_i$ and the mean expected planned processing time $p_i$, the plan start date $s_i$ and plan end date $e_i$ of each production order $i$ at the bottleneck workstation can be calculated as follows:

$$s_i = \max(e_{i-1}, a_i) \qquad (2)$$
$$e_i = s_i + p_i \qquad (3)$$

Here $e_{i-1}$ represents the plan end date of the preceding production order. Thus, the plan start date $s_i$ is determined by either the plan arrival time at the bottleneck workstation $a_i$ or the completion time of the preceding production order, whichever is later. After processing a production order at the bottleneck workstation, the deviation is assessed by comparing the actual end date $\hat{e}_0$ with the plan end date $e_0$ of this completed production order, calculating the deviation $\Delta e = \hat{e}_0 - e_0$. The completed order is designated as index 0, subsequent production orders are systematically labeled with increasing indexes. If the production order is completed ahead of schedule (i.e., $\Delta e < 0$), forward rescheduling is initiated, pushing subsequent orders ahead but not past the earliest plan start date at the bottleneck workstation $a_i$. Conversely, in the case of a delay (i.e., $\Delta e > 0$), backward rescheduling is performed, where any bottleneck idle time gaps between production orders are considered. As illustrated in Equation (2), these gaps arise when the earliest planned start date of the scheduled production order, falls later than the planned end date of the preceding scheduled order. These gaps primarily emerge in periods when the gap between required and available capacity is large. The cumulative gap $G_i$, representing the sum of individual bottleneck idle time gaps $g_k = s_k - e_{k-1}$ preceding production order $i$, is calculated as:

$$G_i = \sum_{k=1}^{i} g_k \qquad (4)$$

For both rescheduling scenarios the adjusted start date $s'_i$ for each production order $i$ has to be calculated, specifically for forward scheduling as follows:

$$s'_i = \max(s_i - \Delta e, a_i) \qquad if\ \Delta e < 0 \qquad (5)$$

Whereas, in case of backward scheduling, at first, the adjusted shift for each production order $\Delta e'_i$, considering idle time gaps at the bottleneck workstation, has to be calculated as:

$$\Delta e'_i = \max(\Delta e - G_i, 0) \qquad if\ \Delta e > 0 \qquad (6)$$

Based on the adjusted shift for each production order $\Delta e'_i$ the adjusted start date $s'_i$ at backward scheduling is then:

$$s'_i = s_i + \Delta e'_i \qquad (7)$$

For both forward and backward rescheduling the adjusted plan end date $e'_i$ for production order $i$ can be calculated based on Equation (3) substituting $s_i$ with the adjusted start date $s'_i$. To summarize, only two planning parameters are essential for scheduling the bottleneck workstation and defining DBR: the CCR-Buffer, and the Shipping-Buffer. The following Table 1 outlines all the notations used.

Table 1. Notations for Forward Scheduling Description.

| Notation | Description |
| --- | --- |
| $I$ | Set of production orders |
| $O(t)$ | Set of production orders considered for scheduling at time $t$ |
| $R(t)$ | Set of released production orders at time $t$ |
| $S$ | Shipping-Buffer |
| $C$ | CCR-Buffer |
| $t$ | Current time |
| $d_i$ | Due date of produciton order $i$ |
| $a_i$ | Earliest plan start date of production order $i$ at bottleneck workstation |
| $s_i$ | Plan start date of production order $i$ at bottleneck workstation |
| $s'_i$ | Adjusted plan start date after rescheduling for production order $i$ |
| $p_i$ | Plan process time of prodcution order $i$ at bottleneck workstation |
| $e'_i$ | Adjusted plan end date after rescheduling for production order $i$ |
| $e_0$ | Plan end date of completed production order at bottleneck workstation |
| $\hat{e}_0$ | workstation |
| $\Delta e$ | Deviation between plan end date and realized end date |
| $\Delta e'_i$ | Adjusted shift for production order $i$ |
| $g_k$ | Size of the $k^{th}$ bottleneck idle time gap |
| $G_i$ | Cumulative bottleneck idle time gap up to production order $i$ |

To improve performance, the original form of DBR dispatches production orders to the bottleneck workstation based on the Earliest-Constraint-Date (ECD), and subsequently on an EDD basis. Performance can be further enhanced by employing more sophisticated or varied combinations of sequencing and dispatching rules, as demonstrated by Thürer and Stevenson [17]. In addition, the incorporation of Buffer Management based on real-time information, further supports production control, as discussed in Woo et al. [18].

### 2.2     Efficient Simulation-Based Parameter Optimization

The concept of SBM is utilized in simheuristics to minimize costs in simulation experiments with limited resources [11]. This approach aims to explore all parameter sets within a predefined solution space efficiently. The SBM method evaluates the quality of simulation iterations by comparing the average costs against a dynamically adjusted percentile of all known iterations, skipping further replications if costs exceed a certain threshold. Key to this method is the initialization of the parameters lower bound (lb), upper bound (ub), percentile step, as well as the criteria for the minimum number of replications and iterations required to start SBM. Seiringer et al. [11] previously successfully applied the SBM concept in an MRP planned production system. Building on this, our current study integrates this SBM with an agent-based discrete event simulation model of a forward-scheduled DBR production system, further demonstrating SBM's broad applicability and effectiveness.

## 3     Simulation Model

In the following section, we initially describe the production system and its modeling within a simulation framework. Subsequently, we detail the conducted numerical study, outlining the tested production system environments, planning parameters, and settings for SBM.

### 3.1     Production System and Simulation Model

In this study, we investigate a multi-item multi-stage make-to-order flow shop production system with limited capacity and stochastics in demand, processing as well as customer-required lead times. Our observed production system is modeled as an agent-based simulation using AnyLogic and consists of five workstations {W1, …, W5}, with workstation four identified as the bottleneck, and three products {P1, P2, P3}. Fig. 1 illustrates the modeled production system, while Fig. 2 highlights the Bill of Materials (BoM), detailing the required workstations for producing the respective components and products. As shown in Fig. 2, the component C2 is initially produced using workstations {W1, W2, W3} along with the raw material R, which is always available and not planned. At the bottleneck workstation W4, component C2 is split into components C1/1 and C1/2. Finally, at workstation W5, component C1/1 is used to manufacture product P1, while C1/2 is utilized to produce either product P2 or P3. Each level of the BoM requires one specific component.

A real-world scenario similar to our observed setup is the automotive assembly line, where engine installation acts as a critical bottleneck due to the complex and time-consuming tasks of fitting the engine and connecting various systems. Such an assembly line is typically organized as a multi-item multi-stage flow shop where vehicles move from one station to the next, each dedicated to a specific part of the assembly process. Additionally, a comparable formulation is found in electronics manufacturing, which also follows a multi-item multi-stage flow shop configuration and faces significant bottlenecks in the assembly of complex multi-component devices such as smartphones.

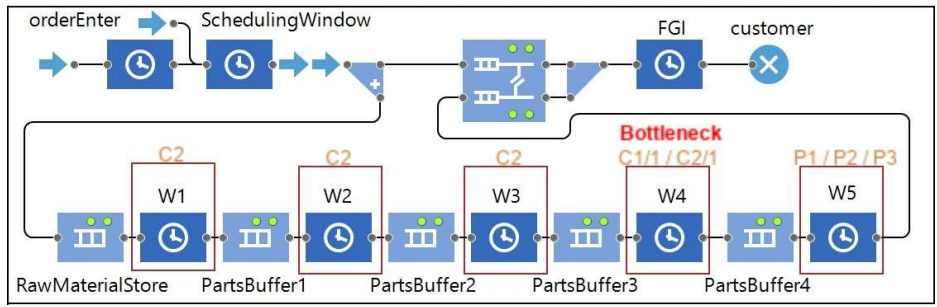

**Fig. 1.** Production System.

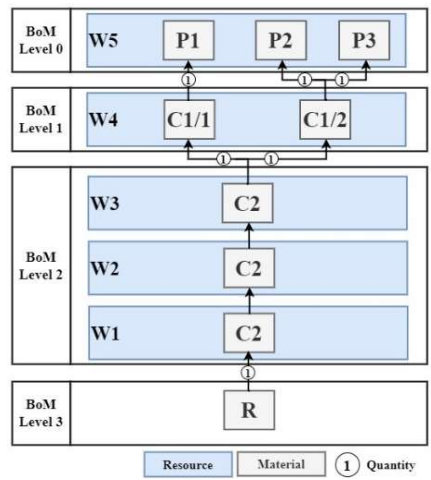

**Fig. 2.** Bill of Materials.

Each customer order is comprised of one three product type, selected through a random categorical distribution with probabilities ∈ {0.25, 0.50, 0.25}, reflecting the varying likelihoods of each product type being chosen. These customer orders have a discrete uniformly distributed lot size within the range of ∈ {1, 2}, from which a single production order is generated. The inter-arrival time of customer orders follows a lognormal distribution with a coefficient of variation (CV) of 0.3. The customer-required due date combines a deterministic part of 6 time units and a variable part that is exponentially distributed, with a mean 16 time units.

As we observe a make-to-order production system, these customer orders are directly transferred to production orders. The expected mean planned processing time at the initial three workstations {W1, W2, W3} is consistently set at 0.65 time units, as only one component, C1, is produced, as shown in Fig. 2. However, the expected mean planned processing time at the bottleneck workstation W4 is 0.70 time units for component C1/1 and 0.75 time units for component C1/2. Moreover, at the final workstation W5, the mean planned processing time is 0.70 time units for product P1, 0.60 time units for product P2, and 0.75 time units for product P3. All these times refer to produce one product respectively component. The available capacity at each workstation is 1440 time units per day. We calculate the shop load by integrating the inter-arrival times, expended mean planned processing times, and constrained capacity. This shop load represents the planned utilization, excluding the stochastic variability introduced by stochastic processing times. To ensure efficiency, we maintain the shop load at each non-bottleneck workstation at 10% lower than at the bottleneck workstation. For example, while a 95% planned utilization is maintained at W4, the other workstations operate at a planned utilization of 85%. These information about applied distributions and parametrization regarding the simulation model can be found in Table 3.

Scheduling for the bottleneck workstation, and consequently the entire production system, is conducted using the forward approach outlined in Section 2.1. During the simulation process at the workstations, the processing time is determined by a lognormal distribution, based on the above stated mean expected planned processing time and coefficient of variation of this expected planned processing time (CV PPT) 0.3, while also accounting for lot size. his CV PPT is varied only at the bottleneck workstation during the numerical study to investigate different behaviors in stochastic production systems. Within the system, production orders are dispatched up to the bottleneck using an Earliest-Constraint-Date (ECD), and post-bottleneck using an EDD, in alignment with the original version of DBR [2]. Finished goods are held in the Finished Goods Inventory (FGI) until reaching the customer-required due date. If tardiness occurs, delivery is executed immediately. Building on the previous description, the distributions were chosen for their non-negativity and widespread application in prior research. These values were calibrated based on preliminary studies to ensure they adequately challenge the production system without causing overload.

### 3.2   Numerical Study

In the following we describe the setup of the experiments for the performed numerical study with our developed simulation model. Using 20 replications per iteration effectively covers a meaningful range of stochastic effects. Without SBM all 20 replications must be simulated. To test how many replications can be saved using SBM for DBR, a first starting point are the evaluated SBM settings described in Seiringer et al. [11]. The settings S1, S2, S3 and S4 and the corresponding values are: with lower bound (lb) and upper bound (ub) combinations of (S1) lb= 0.05, ub= 0.4; (S2) lb = 0.1, ub= 0.5; (S3) lb= 0.02, ub= 0.8; and (S4) lb= 0.02, ub= 0.2. The most stringent setting is (S4), while the least restrictive setting is (S3), a SBM initialization phase of 5 iterations with 20 replications and a minimum number of 3 replications per iteration before SBM is applied to skip a running replication are used.

In addition to SBM, the planning parameter ranges for the DBR model, along with different shop loads and process uncertainties, i.e. CV PPT, for the bottleneck workstation are investigated as detailed in Table 2. Moreover, the applied distributions and parameterization concerning the simulation model are shown in Table 3. At the numerical study the shop load is adapted by changing the inter-arrival time of customer orders. Three different levels (low, medium, high) are investigated for both the shop load and the process uncertainty. This leads to 9 different production system environments. The tested ranges for the CCR-Buffer are from 1 to 12 time units, and for the Shipping-Buffer, they are from 1 to 24 time units, each with a step size of 1 time unit. Both ranges were set based on preliminary studies. In total 288 different parameter combinations are tested for DBR without SBM (no/SBM) and DBR in combination with S1, S2, S3 and S4 for each of the 9 production system environments. This leads to 5760 individual replications for no/SBM at each production system environment and also potentially 5760 for each SMB setting {S1, ..., S4}. If SBM is enabled for the DBR model the 5760 should not be consumed to find the optimal solution. Hence, in total 259 200 potential individual replications are examined (9 production system environments x 5760 individual replications x 5 methods). Each replication is run for 8760 hours simulation time (365 days) with a warm-up phase of 760 hours (32 days). Each experiment is run subsequently on the same hardware a computer with an Intel i5-10500 CPU (3.1Ghz) and 32 Gb RAM. The parameter combinations are generated using R, then exported to a H2 database and subsequently queried by the Anylogic DBR model. The results are afterwards also stored into the same H2 database.

**Table 2.** DBR Parameter Ranges.

| | Min | Max | Step Size | Iterations |
|---|---|---|---|---|
| Shop Load | 0.85 | 0.95 | 0.05 | 3 |
| CV PPT | 0.30 | 0.90 | 0.30 | 3 |
| Different production system environments | | | | 9 |
| CCR-Buffer $C$ [TU] | 2 | 12 | 1 | 12 |
| Shipping-Buffer $S$ [TU] | 2 | 24 | 1 | 24 |
| Total Iterations per production system environment (no/SBM) | | | | 288 |
| Individual Replications per production system environment (no/SBM) | | | | 5760 |

**Table 3.** Applied Distribution and Parametrization.

| | Distribution Type | Parametrization |
|---|---|---|
| Product Type | Categorical | {P1 = 0.25, P2 = 0.50, P3 = 0.25} |
| Size of Customer Order | Discrete Uniform | U(1,2) |
| Inter-Arrival Time Customer Orders | Lognormal | Log($\mu$ = cal. based on shop load {0.85, 0.90, 0.95}, $\sigma$ = 0.3) |
| Customer-Required Due Date Fix Part | Deterministic | 6 |
| Customer-Required Due Date Variable Part | Exponential | Exp($\lambda$ = 1/16) |
| Planned Processing Time {W1, W2, W3} | Lognormal | Log($\mu$ = 0.65, $\sigma$ = 0.3) |
| Planned Processing Time W4 | Lognormal | Log($\mu$ = {C1/1 = 0.70, C1/2 = 0.75}, $\sigma$ = {0.3, 0.6, 0.9}) |
| Planned Processing Time W5 | Lognormal | Log($\mu$ = {P1 = 0.70, P2 = 0.60, P3 = 0.75}, $\sigma$ = 0.3) |

## 4      Simulation Results and Discussion

In this section, we present the findings from our simulation study and address the research questions RQ1 to RQ4. To analyze the production system performance, overall costs are selected, which encompass the combined expenses of Work-in-Process (WIP) costs, FGI costs, and tardiness costs. The costing value for WIP is set at 0.5, half of the FGI costing value, which is 1. This distinction in costing values is logical, as storing finished goods typically incurs higher costs than holding WIP. The tardiness costs are set to 19, which represent a target service level of 95%. A service level of 0.95 indicates a 95% likelihood that the available inventory will meet customer demand, thus avoiding any stockouts. First, our analysis explores the impact of varying shop loads and process uncertainties on overall costs and parameterization. Subsequently, we examine the potential for reducing simulation budgets through the application of SBM.

### 4.1      Effects of Different Shop Load and Process Uncertainty

Fig. 3 illustrates the minimum overall costs per time unit, expressed in cost units (CU) for the production system at different shop loads of 0.85, 0.90 and 0.95 and CV PTT at the bottleneck workstation W4, which is show in Fig. 1.

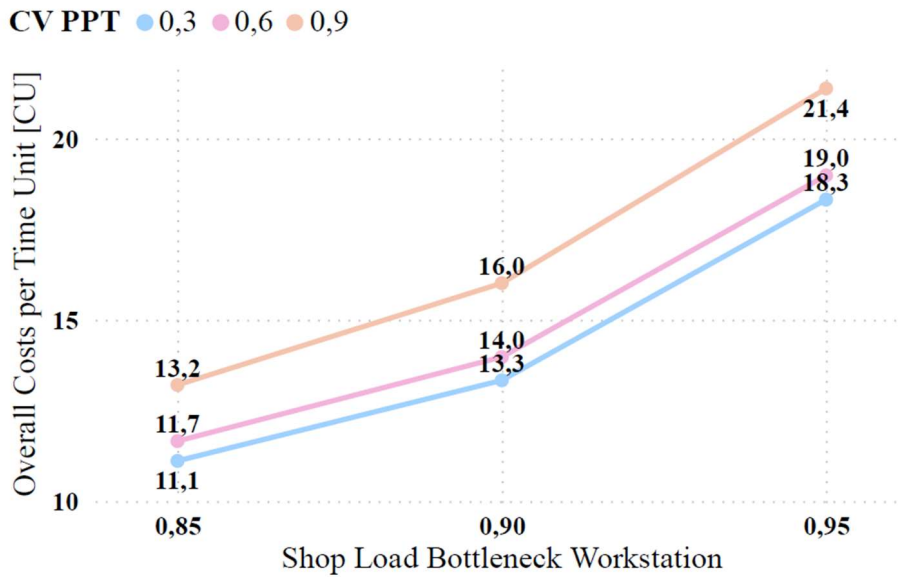

**Fig. 3.** Min. Overall Costs per Planned Utilization & CV PPT.

An increase in both shop load and process uncertainty, i.e. CV PPT, results in higher overall costs. This rise in costs can be attributed to two main factors. Firstly, at higher levels of shop load, the bottleneck workstation experiences fewer idle periods, thus limiting its ability to compensate for delays. Secondly, higher processing uncertainty results in larger discrepancies between the planned and actual end dates at the bottleneck workstation, which leads to larger rescheduling as indicated by Equation (4) and Equation (8). To discuss RQ1 in Table 3 the DBR parameterizations of the identified minimum overall costs, are presented for a comprehensive analysis.

Table 3. Optimal DBR Planning Parameters.

|  | Shop Load | | |
|---|---|---|---|
|  | 0.85 | 0.9 | 0.95 |
| CV PPT 0.30 | C = 6<br>S = 7 | C = 7<br>S = 8 | C = 8<br>S = 15 |
| CV PPT 0.60 | C = 6<br>S = 8 | C = 7<br>S = 9 | C = 8<br>S = 15 |
| CV PPT 0.90 | C = 6<br>S = 9 | C = 7<br>S = 11 | C = 8<br>S = 19 |

*all values expressed in Time Units [TU]*

As demonstrated in Table 3, variations in the production system environment have a marginal effect on the CCR-Buffer, with only a slightly longer buffer required at higher shop load. This necessity is because higher shop loads at the bottleneck workstation also increase the utilization of the preceding workstations, given that the inter-arrival time is adjusted in our model. Consequently, this increased utilization requires a longer CCR-Buffer to prevent the bottleneck from starvation. Whereas, a slight increase in process uncertainty at the bottleneck workstation necessitates a longer Shipping-Buffer. This buffer extension compensates for the growing variances between planned and actual completion times. Moreover, an increase in either of the two buffers enlarges the scheduling window, as specified in Equation (1). This enlargement enhances the scheduling flexibility by considering a broader set of production orders for production release.

### 4.2   Simulation Budget Reduction Potential Through SBM

The intention of SBM is, to safe simulation time by skipping not promising replications. Skipping not promising solutions should not lead to worser solution quality. The saved simulation budget can then, for example, be used to test additional parameter combinations. In our DBR investigations the best solution is represented by the minimum overall costs within the selected result subset. To validate the effectiveness of SBM within the simulation model, it is essential to demonstrate that the minimum overall costs identified by SBM are equivalent to those found in the model without SBM. The data shown in Table 4 demonstrates the ability of SBM to identify the same minimum overall costs as those found by a full factorial (FF) enumeration for each production system environment. Since identical costs are observed for each SBM setting {S1, ..., S4}, only a single representative value is displayed for SBM.

**Table 4.** Min. Overall costs for no/SBM and SBM settings.

|  |  | Shop Load | | |
|---|---|---|---|---|
|  |  | 0.85 | 0.9 | 0.95 |
| CV PPT | 0.30 | FF = 11.1138<br>SBM = 11.1138 | FF = 13.3347<br>SBM = 13.3347 | FF = 18.3174<br>SBM = 18.3174 |
|  | 0.60 | FF = 11.6603<br>SBM = 11.6603 | FF = 13.9776<br>SBM = 13.9776 | FF = 18.9818<br>SBM = 18.9818 |
|  | 0.90 | FF = 13.2072<br>SBM = 13.2072 | FF = 16.0121<br>SBM = 16.0121 | FF = 21.3783<br>SBM = 21.3783 |

*all values expressed in Costs per Time Unit [CU]

As seen in Table 4 the minimum overall costs could be found at each SBM setting for each production system environment. Pertaining to RQ3, this indicates that the likelihood of missing an optimal solution when using SBM, as opposed to conducting a full factorial enumeration, is negligible. With the established accuracy of the SBM implementation, we can analyze its performance relative to a full factorial enumeration using the data in following 5. This Table 5 details the number of replications used, with the improvement delta between no/SBM and with SBM application indicated in brackets. For instance, the 5760 replications for the no/SBM correspond to 288 parameters, each with 20 replications. Δ1 denotes the relative difference in the number of replications between the no/SBM and SBM setting S1 to S4, calculated as (SBM replications – no/SBM replications) divided by no/SBM replications. Δ2 represents the average improvement delta across all four SBM settings for the identical shop load but different process uncertainty. Whereas, Δ3 indicates the average improvement delta for the same SBM setting across all three levels of process uncertainty. At every SBM setting, a significant performance advantage is evident compared to the setting without SBM (no/SBM). However, as either shop load or process uncertainty increases, the advantage in terms of replication savings lessens, which is observable when comparing Δ2 across identical shop loads and identical process uncertainties. This trend of reduced replication savings under SBM settings with increased shop loads or process uncertainty can be linked to the fact that higher shop loads induce more pronounced variations between replications. This leads to a greater probability of experiencing elevated costs and consequently requires the continuation of the current iteration without skipping. Across all SBM settings, setting S4, which is the most stringent, realizes the most substantial savings. S4 uniformly surpasses the other settings (S1, S2, and S3) within the same production system environments and maintains the superior average for varying process uncertainty Δ3, demonstrating the optimal delta values of -0.68, -0.61, and -0.47. These impressive results demonstrate that SBM with S4 not only yields the same minimum overall costs but also significantly reduces simulation budget.

Table 5. SBM performance regarding replications.

| | | **Shop Load 85%** | | | | |
|---|---|---|---|---|---|---|
| | | w/oSBM | S1 (Δ1) | S2 (Δ1) | S3 (Δ1) | S4 (Δ1) | Avg(Δ2) |
| CV PPT | 0.3 | 5760 | 1991 (-0,65) | 2279 (-0,6) | 2716 (-0,53) | 1690 (-0,71) | -0.62 |
| | 0.6 | 5760 | 2096 (-0,64) | 2391 (-0,58) | 2804 (-0,51) | 1762 (-0,69) | -0.61 |
| | 0.9 | 5760 | 2443 (-0,58) | 2730 (-0,53) | 3061 (-0,47) | 2102 (-0,64) | -0.56 |
| | Avg(Δ3) | | -0.62 | -0.57 | -0.5 | -0.68 | |
| | | **Shop Load 90%** | | | | |
| | | w/oSBM | S1 (Δ1) | S2 (Δ1) | S3 (Δ1) | S4 (Δ1) | Avg(Δ2) |
| CV PPT | 0.3 | 5760 | 2307 (-0,6) | 2587 (-0,55) | 2913 (-0,49) | 1952 (-0,66) | -0.58 |
| | 0.6 | 5760 | 2494 (-0,57) | 2766 (-0,52) | 3046 (-0,47) | 2129 (-0,63) | -0.55 |
| | 0.9 | 5760 | 2950 (-0,49) | 3189 (-0,45) | 3349 (-0,42) | 2617 (-0,55) | -0.48 |
| | Avg(Δ3) | | -0.55 | -0.51 | -0.46 | -0.61 | |
| | | **Shop Load 95%** | | | | |
| | | w/oSBM | S1 (Δ1) | S2 (Δ1) | S3 (Δ1) | S4 (Δ1) | Avg(Δ2) |
| CV PPT | 0.3 | 5760 | 3202 (-0,44) | 3401 (-0,41) | 3453 (-0,4) | 2890 (-0,5) | -0.44 |
| | 0.6 | 5760 | 3309 (-0,43) | 3501 (-0,39) | 3538 (-0,39) | 3019 (-0,48) | -0.42 |
| | 0.9 | 5760 | 3458 (-0,4) | 3595 (-0,38) | 3640 (-0,37) | 3204 (-0,44) | -0.40 |
| | Avg(Δ3) | | -0.42 | -0.39 | -0.39 | -0.47 | |

## 5     Conclusion

In this publication, we integrated SBM into the production planning and control method DBR to reduce simulation budget for method parameterization. Moreover, our analysis examined the impact of different production system environments, i.e., varying shop loads and process uncertainty, on both the performance and parameterization of DBR and the efficacy of SBM. At our numerical study we simulated a multi-stage multi-item DBR planned and controlled production system with limited capacity. Findings reveal that higher shop load or process uncertainty increases overall costs and the Shipping-Buffer, due to growing discrepancies in planned versus actual completion times. However, only at higher shop load a slight increase of the CCR-Buffer is required. Additionally, increasing either buffer widens the scheduling window, enhancing production system flexibility. SBM consistently matched the minimum overall costs of a full factorial enumeration across all SBM settings and production system environments, also offering notable simulation budget savings. Yet, these savings lessen with increased shop load or process uncertainty due to rising cost variations, prompting continued iteration. The most substantial simulation budget reduction was noted with the strictest SBM setting, independent of the production system environment. Future research on DBR should explore more complex production systems, with varied buffers for divergent production paths. For SBM, advancements should aim at refining the algorithm's precision by identifying parameters with significant performance impact during runtime and adjusting the step size of them dynamically, i.e., smaller in promising areas of the search space and larger otherwise.

## Acknowledgments

This research was funded in whole or in part by the Austrian Science Fund (FWF) [P32954-G] G and the Funding Agency of the State of Upper Austria with the project FO999905125. For open access purposes, the author has applied a CC BY public copyright license to any author accepted manuscript version arising from this submission.